\documentstyle [fleqn,12pt] {article}
\begin{document}

   \begin{titlepage}

 \begin{center}

   {\huge STRUCTURE FORMATION  \vspace{0.2cm}\\
    IN THE UNIVERSE FROM  \vspace{0.2cm}\\
	TEXTURE INDUCED \vspace{0.2cm}\\
	FLUCTUATIONS}
    \vspace{0.2cm}  \\     {\bf \large Ruth Durrer and Zhi--Hong Zhou}
    \vspace{1cm} {\large
   \\ Universit\"at Z\"urich, Institut f\"ur Theoretische Physik,
      Winterthurerstrasse ~190, \\ CH-8057 Z\"urich, Switzerland}
        \vspace{2cm}

 \end{center}

\begin{abstract}
The topic of this letter is structure formation with topological
defects. We first present a partially new, fully local and gauge
invariant system of perturbation equations to treat microwave background
and dark matter fluctuations induced by topological defects (or any
other type of seeds). We show that this treatment is  extremly well
suited for  linear numerical analysis  of structure formation by
applying it to the texture scenario. Our numerical results cover a
 larger dynamical range than previous investigations and are
complementary since we use substantially different methods.
\end{abstract}
 \vspace{2cm}


 \end{titlepage}

Despite great effort and considerable progress, the problem of structure
formation in the universe remains basically unsolved. Observations show that
the fluctuation spectrum on the large scales observed by COBE should be
not very far from scale invariant \cite{Sm,SV,Ben}. This has been
considered as great success for inflationary models which predict a scale
invariant fluctuation spectrum. In this letter we consider an alternative
class of models which also yield a scale invariant spectrum of Cosmic
Microwave Background (CMB) fluctuations: Models where
perturbations are seeded by global topological defects which can form
during symmetry breaking phase transitions in the early universe \cite{Ki}.
To be specific,
we consider texture, $\pi_3$--defects which lead to event singularities
in four dimensional spacetime \cite{Tu}. A common feature of global
topological defects is the behaviour of the energy density in the scalar
field which scales like
 $\rho_T \propto 1/(at)^{2}$ and thus represents always the same fraction
of the total energy density of the universe.
\begin{equation} \rho_T/\rho \sim 8\pi G\eta^2 \equiv 2\epsilon ~,
\end{equation}
where $\eta$ determines  the symmetry breaking scale.
The background spacetime is a Friedmann--Lema\^{\i}tre universe with
$\Omega=1$. We choose conformal coordinates  such that
\[ ds^2 = a^2(-dt^2 + \delta_{ij}dx^idx^j)  ~. \]
Numerical analyses of CMB fluctuations from topological defects on large
scales  have been
performed in \cite{BR,PST}. A spherically  symmetric approximation is
discussed in \cite{DHZ}. Results for intermediate scales
are presented in \cite{CFGT}.
All these investigations (except the quite rough spherically symmetric
calculation \cite{DHZ}) use linear cosmological perturbation theory in
synchronous gauge and take into account only scalar perturbations. In
this letter we derive a fully gauge invariant and local system of
perturbation equations. The (non local) split into
scalar, vector and tensor modes on the hypersurfaces of constant time
is not performed. We  solve the equations numerically in a cold dark
matter (CDM) universe with global texture. In this letter, we
present the main results. In a longer paper we give detailed
derivations of the equations and fully describe  our numerical
methods \cite{DZ}.  Since there are
no spurious gauge modes in our initial conditions, their is no danger
that these may grow in time and much of the difficulties to choose correct
initial conditions (see e.g. \cite{PST}) are removed.

We find that previously ignored vector and tensor fluctuations
contribute approximately 25 \% to the total microwave background
anisotropies (see Fig.~1).

We calculate the microwave background anisotropies on angular scales
which are larger than the angle which subtends the horizon scale at
decoupling of matter and radiation, $\theta> \theta_d$. For $\Omega =1$
and $z_d \approx 1000$
\begin{equation} \theta_d = 1/\sqrt{z_d+1} \approx 0.03 \approx 2^o
	~. \end{equation}
It is therefore sufficient to study the generation and evolution
of microwave background fluctuations after recombination. During this
period, the photons are decoupled from baryonic matter and
 free stream, influenced solely by cosmic gravitational redshift and by
perturbations in the gravitational field (if the medium is not reionized).
The photon distribution function which lives in seven dimensional
relativistic
phase space $P_0{\cal M}=\{(x,p)\in T{\cal M} | g(x)(p,p)=0 \}$, obeys
Liouville's equation
\begin{equation} X_g(f)=0    \label{L}~.\end{equation}
In a tetrad basis $e_\mu$,  $X_g$ is given by (see e.g. \cite{Ste})

\begin{equation}
	X_g = (p^\mu e_\mu + \omega^i_{\:\mu}(p)p^\mu{\partial
	\over \partial p^i})
  ~, \label{liou}	\end{equation}
where $\omega^\nu_{\:\mu}$ are the connection 1--forms of
$({\cal M},g)$ in the basis $e^\mu$.

The metric of a perturbed Friedmann universe with $\Omega=1$
 is given by $ds^2=g_{\mu\nu} dx^\mu dx^\nu$ with

\begin{equation} g_{\mu\nu} = a^2(\eta_{\mu\nu} + h_{\mu\nu}) =
	a^2 \tilde{g}_{\mu\nu}
  ~, \end{equation}
where $(\eta_{\mu\nu}) = diag(-,+,+,+)$ is the flat Minkowski metric and
$|h_{\mu\nu}|\ll1$ is a small perturbation. We now use the fact that the
motion of photons is conformally invariant. Taking into account the
different affine parameters, (\ref{L}) is equivalent to

\begin{equation} (X_{\tilde{g}}f)(x,ap)=0  ~.
	\label{conform}		\end{equation}
If $\bar{e}^\mu$ is a tetrad in Minkowski space,
$e_\mu = \bar{e}_\mu + (1/2)h_\mu^\nu\bar{e}_\nu$ is a tetrad w.r.t the
perturbed geometry $\tilde{g}$. For $(x,\bar{e}_\mu p^\mu)\in \bar{P}_0$,
thus,
$(x,e_\mu p^\mu)\in P_0$. We can therefore define the perturbation of
the distribution function $F$ by
\begin{equation}
   f(x,p^\mu e_\mu) = \bar{f}(x,p^\mu \bar{e}_\mu) + F(x,p^\mu\bar{e}_\mu)
  ~. \end{equation}
Furthermore, we set $p^i=p\gamma^i$, with $p^2=\sum_{i=1}^3(p^i)^2$
and $v=ap$.
Liouville's equation for $f$ then yields a perturbation equation for $F$.
We choose the natural tetrad $e_\mu=\partial_\mu +
	(1/2)h_\mu^\nu\partial_\nu$.
Using (\ref{L},{\ref{liou}) and (\ref{conform}) we obtain
\begin{equation}
(\partial_t +\gamma^i\partial_i)F = -[\dot{H}_L +(A,_i +
	{1\over 2}\dot{B}_i)\gamma^i +
	(\dot{H}_{ij}-B_{i,j})\gamma^i\gamma^j]v{df\over dv} ~,
 \label {LF} \end{equation}
where we have parametrized
\begin{equation} \left(h_{\mu\nu}\right) = \left(\begin{array}{ll}
		2A & 2B_i \\
                2B_i & 2H_L\delta_{ij}
                                +2H_{ij} \end{array}\right),
 \label{scalar} \end{equation}
with $H_i^i=0$.

We now define
\[ m = (1/4){4\pi\over \rho_ra^4}\int Fv^3dv ~.\]
$4m$ is the fractional perturbation of the brightness $\iota$,
\[ \iota = a^{-4} \int f v^3dv ~. \]
Setting $\iota = \bar{\iota}(T(\gamma,x))$, one finds that $m$
corresponds to the fractional perturbation in the temperature,
\begin{equation}
	T(\gamma,x) = \bar{T}(1+m(\gamma,x)) ~.\label{T} \end{equation}
A more explicit derivation of (\ref{T}) is given in \cite{d94}).
Integrating (\ref{LF}) $v^3dv$,  we obtain

\begin{equation}
	\partial_tm+\gamma^i\partial_im= \dot{H}_L +(A,_i +
	{1\over 2}\dot{B}_i)\gamma^i +
	(\dot{H}_{ij}-B_i,_j)\gamma^i\gamma^j ~.
\label{Lm}
\end{equation}

It is well known that the equation of motion for photons only couples to
the Weyl part of the curvature (null geodesics are conformally invariant).
The r.h.s. of (\ref{Lm}) is given by first derivatives of the metric only
which could at most represent integrals of the Weyl tensor. To obtain
a local, non integral equation, we thus rewrite (\ref{Lm}) in terms of
$\triangle m$.  In fact, defining

\[ \chi = \triangle m - (\triangle H_L-{1\over 2}H,_{ij}^{ij}) -
	{1\over 2}\triangle B_i\gamma^i  ~, \]
(\ref{Lm}) yields for $\chi$ the equation of motion

\begin{equation}
 (\partial_t +\gamma^i\partial_i)\chi = -3\gamma^i\partial^jE_{ij}
	- \gamma^k\gamma^j\epsilon_{klj}
	\partial_lB_{ij} ~ ,
  \label{Lchi}  \end{equation}
where $\epsilon_{kli}$ is the totally antisymmetric tensor in three
dimensions,
$E_{ij}$ and $B_{ij}$ are the electric and magnetic part of the Weyl tensor.
The spatial indices in this equation are raises and lowered with
	$\delta_{ij}$
and therefore no care is taken in the index position. Double indices are
summed over irrespective of their position.

In eqn. (\ref{Lchi}) the contribution from the electric part of the
Weyl tensor does not contain tensor perturbations.  On the other hand,
scalar perturbations do not induce a magnetic gravitational field. The
second contribution to the source term in (\ref{Lchi}) represents a
combination of vector and tensor perturbations. If vector perturbations
are negligible, the two terms on the r.h.s of (\ref{Lchi}) therefore
represent a split into scalar and tensor  perturbations which is local.

In terms of  metric perturbations, the electric and magnetic part of the
Weyl tensor are given by (see, e.g. \cite{Ma})
\begin{eqnarray}
 E_{ij} &=&  {1\over 2}[\triangle_{ij}(A-H_L) -\dot{\sigma}_{ij}
		-(\triangle H_{ij}+{2\over 3}H_{lm}^{,lm}\delta_{ij})
	- H_{il}^{,l},_j- H_{jl}^{,l},_i] \label{E} \\
 B_{ij} &=& {1\over 2}(\epsilon_{ilm}\sigma_{jm},_l +
	\epsilon_{jlm}\sigma_{im},_l ) ~,
  \label{B}  \end{eqnarray}
\[ \mbox { with }~~ \sigma_{ij}= {1\over 2}(B_i,_j+B_j,_i)-
	{1\over 3}\delta_{ij}B_l^{,l} -\dot{H}_{ij} ~~~
  \mbox { and }~~ \triangle_{ij} =\partial_i\partial_j
-(1/3)\delta_{ij}\triangle ~.\]

Since the Weyl tensor of Friedmann Lema\^{\i}tre universes vanishes, the
rhs of (\ref{Lchi}) is manifestly gauge invariant. (This is the so called
Stewart lemma \cite{SW}.) Therefore also the variable
$\chi$ is gauge invariant.

The general solution to (\ref{Lchi}) is given by

\begin{equation}
	\chi(t,\mbox{\boldmath{$x,\gamma$}}) =
	\int_{t_i}^t S_T(t',\mbox{\boldmath{$x$}}+(t'-t)
   \mbox{\boldmath{$\gamma,\gamma$}})dt'
	~ + ~ \chi(t_i,\mbox{\boldmath{$x$}}+(t_i-t)
	\mbox{\boldmath{$\gamma,\gamma$}}) ~,
\label{chi} \end{equation}
where $S_T$ is the source term given on the rhs of (\ref{Lchi}).

The electric and magnetic part of the Weyl tensor are determined by
the perturbations in the energy momentum tensor via Einstein's
equations.
We assume that the source for the geometric perturbations is given by
the scalar field and  dark matter. The contributions from radiation
may be neglected. Furthermore, vector perturbations of the dark matter
(which decay quickly) are neglected.
The  divergence of $E_{ij}$ is then determined  by
(see, \cite{Ma} or \cite{BDE})

\begin{equation} 	\partial^jE_{ij} = -8\pi G\rho_{DM}\gamma^iD_i -
	8\pi G(\partial_i\delta T_{00} +3({\dot{a}\over a})\delta T_{0i})
	+ 12\pi G\partial^j\tau_{ij}
	 \label{dE} ~,
\end{equation}
where
\[\tau_{ij} \equiv T_{ij} -(a^2/3)\delta_{ij}T^l_l =
		\tau_{ij}^{(texture)}
  = \phi,_i\phi,_j -(1/3)\delta_{ij}(\nabla\phi)^2 ~,\]
\[ \delta T_{0j}= \delta T_{0j}^{(texture)} = \dot{\phi}\phi,_j ~~, \]
\[ \delta T_{00}= \delta T_{00}^{(texture)} = {1\over 2}((\dot{\phi})^2
	 + (\nabla\phi)^2) ~,\]
and $D_j$ is a gauge invariant perturbation variable  for the density
gradient (see \cite{Ma,BDE,DZ}). For scalar perturbations $ D_j =
	\partial_jD$.
The evolution equation for the dark matter density perturbation
is given by
\begin{equation}
   \ddot{D} + ({\dot{a}\over a})\dot{D} - 4\pi Ga^2\rho_{DM}D
	= 8\pi G \dot{\phi}^2
\label{dm} ~. \end{equation}

The equation for $B_{ij}$ is  more involved. A somewhat cumbersome
derivation \cite{DZ} yields

\begin{equation}
	\ddot{B}_{ij} + 3({\dot{a}\over a}) \dot{B}_{ij} -\triangle B_{ij}
	= 8\pi G{\cal S}^{(B)}_{ij} ~, \label{Bij}
\end{equation}
\[ \mbox{with }~~~ {\cal S}^{(B)}_{ij}= \epsilon_{lm(i}\delta T_{0l},_{j)m}
	- (\dot{a}/a)\epsilon_{lm(i}\tau_{j)l},_m  ~.\]
Here $(i...j)$ denotes symmetrization in the indices $i$ and $j$.

To these equations we have to add the evolution equation of the scalar
field,

\begin{equation} \ddot{\phi} +2(\dot{a}/a)\dot{\phi} -\triangle\phi =
	a^2\lambda\phi(\phi^2-\eta^2)   ~. \label{phi}
\end{equation}

We have solved the closed hyperbolic system (\ref{Lchi}, \ref{dE},
 \ref{dm}, \ref{Bij}, and \ref{phi}) numerically on a $192^3$ grid for
different initial conditions  on a NEC--SX3 computer at
the Centro Svizzero di Calcolo Scientifico (CSCS). The numerical
methods employed and
 the different tests of our programs are described in \cite{DZ}.
Here we just want to present the main results.

Since on subhorizon scales gauge dependent and gauge invariant variables
do not differ substantially we can interpret the  variables $D$ and
$\chi$ by
\[  D = (\delta\rho/\bar{\rho}) = \delta ~~ \mbox{ and }~~~~
    \chi = \triangle (\delta T/\bar{T})  ~. \]
Using fast Fourier transforms we calculate the spectrum
$P(k) = |\delta(k)|^2$ and $\delta T/\bar{T}$, which we then  expand
in spherical harmonics
\begin{equation}  (\delta T/\bar{T})(t_0,\mbox{\boldmath{$x,\gamma$}}) =
 \sum_{lm}a_{lm}(x)Y_{lm}(\mbox{\boldmath{$\gamma$}})
 ~. \end{equation}
As usual we  assume that the average over different observer
positions coincides with the ensemble average and determine
\begin{equation} c_l = {1\over (2l+1)N_x} \sum_{m,x}
	|a_{lm}(\mbox{\boldmath{$x$}})|^2
 \label{cl} ~. \end{equation}
We have performed  10
 simulations on a $192^3$ grid with about 100 different observer
positions for each simulation. The average harmonic amplitudes with
$1\sigma$ variance are shown in Fig.~2.
The low order multipoles depend strongly on  the random initial
conditions (cosmic variance), like in the spherically symmetric
simulation \cite{DHZ}. From Fig.~2 it
is clear that the texture scenario is compatible with a scale
invariant spectrum. The main difference of the currently favored
scenarios with inflation induced perturbations lies in the distribution
of fluctuations which is non Gaussian in models with topological defects.
The pixel distribution for $\Delta T/T$ in the sky is  negatively
skewed. Our quadrupole apmlitude is given by
\[ Q = (0.53 \pm 0.16)\epsilon \]
To reproduce the COBE amplitude $ Q_{COBE}= (0.6 \pm 0.1)10^{-5}$
\cite{Ben},
we have to normalize our spectrum by choosing the phase transition
scale $\eta$
\begin{equation} \epsilon = 4\pi G\eta^2 = (1.1 \pm 0.5)10^{-5}
	\label{ep}~. \end{equation}
This value is comparable with the value of $\epsilon$ obtained in
\cite{BR,PST}.

The power spectrum of dark matter density fluctuations
is shown in Fig.~3. To be compatible with
observations $(\sigma_8\approx 1)$, we have to introduce a somewhat
high bias factor of
\[ b \approx  4\pm 2 ~. \]
(The bias factor takes into account that the observed clustering of light
does not necessarily coincide with the clustering of the underlining
dark matter distribution.)
Observations
and simulations of nonlinear clustering of dark matter and baryons
\cite{CO} hint that a bias factor b=2 -- 2.5 might be reasonable.

In this letter we have presented a closed system of cosmological
 perturbation equations which are not plagued by gauge modes and which
is well suited for numerical analysis. Our numerical results are
consistent with previous investigations \cite{BR,PST,DHZ}, indicating
that the texture scenario of structure formation yields somewhat too
much power on small scales.
\vspace{2cm}

{\large\bf Acknowledgement}
We thank the staff at CSCS, for valuable
support. Especially we want to mention Andrea Bernasconi, Djiordic
Maric and Urs Meier.

\newpage

{\Large Figure Captions}
\vspace{1cm}\\
{\bf Fig.~1}\\
The amplitude of the electric and magnetic source terms to the photon
equation if motion are shown as a function of wavenumber $k$ in arbitrary
scale. On very large scales the magnetic part contributes about
1/4 decaying to roughly
1/10 on small scales ($\bar{E}= {1\over 3}\sum_i(\partial^jE_{ij})^2 ~,
{}~~ \bar{B} ={1\over 6}\sum_{ij}(\epsilon_{jlk}\partial^lB_{ki})^2$).
\vspace{1cm}\\
{\bf Fig.~2}\\
The harmonic amplitudes $l(l+1)c_l$ are shown with $1\sigma$ error bars.
The slight rise at large $l$ is due to the finite size of the texture
core. No smoothing is applied.
\vspace{1cm}\\
{\bf Fig.~3}\\
The power spectrum of the CDM fluctuations induced by texture.
(The vertical scale is arbitrary.)

\end{document}